# Existence Conditions for Phononic Frequency Combs

Zhen Qi[1], Curtis R. Menyuk[1], Jason J. Gorman[2], and Adarsh Ganesan[2*]

[1]Department of Computer Science and Electrical Engineering, University of Maryland, Baltimore County, Baltimore, MD 21250, USA

[2]National Institute of Standards and Technology, Gaithersburg, MD 20899, USA

[*]Contact: adarshvganesan@gmail.com

**The mechanical analog of optical frequency combs, phononic frequency combs, has recently been demonstrated in mechanical resonators and has been attributed to coupling between multiple phonon modes. This paper investigates the influence of mode structure on comb generation using a model of two nonlinearly coupled phonon modes. The model predicts that there is only one region within the amplitude-frequency space where combs exist, and this region is a subset of the Arnold tongue that describes a 2:1 autoparametric resonance between the two modes. In addition, the location and shape of the comb region are analytically defined by the resonance frequencies, quality factors, mode coupling strength, and detuning of the driving force frequency from the mechanical resonances, providing clear conditions for comb generation. These results enable comb structure engineering for applications in areas as broad as sensing, communications, quantum information science, material science, and molecular science.**

Optical frequency combs have received considerable interest due to the stable broadband comb structure that can be generated, which has been a powerful tool in many applications, including optical clocks, spectroscopy, and microwave frequency synthesis [1,2]. Like optical resonators, mechanical resonators have also been shown to be capable of generating equally spaced vibrational frequencies due to mechanical mixing and mode coupling [3-14]. Early demonstrations [3-6] revealed that by electrically driving coupled mechanical resonators or multiple modes in a single resonator using multiple drive frequencies simultaneously, a comb-like structure can be generated in the frequency domain. More recently, it was shown that a phononic frequency comb with well-defined frequency structure can be generated with a single mechanical resonator that is driven with a single frequency [7]. In this case, length extensional and flexural vibration modes are coupled through mechanical nonlinearities, providing a mechanism for mode coupling that generates phononic frequency combs when the amplitudes of the coupled modes saturate. Additional experimental observations of phononic frequency combs with a single drive frequency have since been reported that support the results in [7], including comb generation in a nanomechanical beam resonator [8], a coupled translational-rotational resonator [11], and a membrane resonator [12]. The parametric mode coupling seen in [7-16] provides a path for engineering the comb structure and will likely find applications in sensing, communications, and quantum information science, similar to optical frequency combs. In addition, this new phenomenon of phononic combs could be exploited in material and molecular sciences, for instance in the investigation of nonlinear phononics [17].

Despite the growing number of experimental observations of phononic frequency combs in mechanical resonators, it is still largely unclear how the resonance frequencies and quality factors of the interacting modes influence the generation and properties of the comb. The Fermi-Pasta-Ulam framework has previously been used to prove that a comb can be generated with a single drive frequency for a mechanical coupled-mode system [18]. This analysis presented time-domain results that show that the comb can be phase coherent and that generation can be achieved through a wide range of nonlinearities and the number of coupled modes. More recently, a nonlinear friction mechanism has been shown analytically to be capable of generating a comb using just a single vibrational mode in a nanomechanical resonator [19]. In this paper, we analyze the effect of the resonance frequencies and quality factors of the coupled modes on the amplitude-frequency behavior of the comb. We apply the slowly varying envelope approximation to two coupled mechanical modes with a 2:1 autoparametric resonance and derive analytical existence conditions for phononic frequency combs. Using the derived existence conditions, the position and shape of the comb region relative to resonance are also presented. This analysis provides guidelines for tailoring phononic frequency combs in mechanical resonators.

The generation of phononic frequency combs in the mechanical resonator described in [7,9,10,13] and shown schematically in Fig. 1a involves two steps. First, mode 1, which is a length extensional vibration mode, excites mode 2, which is a flexural vibration mode, when increasing the drive amplitude $F$ above a certain threshold (see Fig. 1b). Second, mode 2 feeds back into mode 1 until the amplitudes of both modes oscillate and there is a continuous exchange of energy between the two modes. This is similar to a phenomenon found in optical parametric oscillators (OPO) where the idler is generated from the pump by increasing the pump power over a threshold [20,21]. When further increasing $F$, modes 1 and 2 experience a temporal oscillation, similar to the slow time scale in Kerr combs. This corresponds to a Hopf bifurcation of two eigenmodes that are symmetric to the real axis. These eigenmodes transition from stable to unstable for specific drive, or pump, conditions, resulting in phononic frequency combs. This behavior is described by two coupled phonon modes with quadratic coupling nonlinearities and a 2:1 autoparametric resonance. The coupling is a result of a nonlinear strain relationship between the length-extensional and flexural modes [22], and the equations of motion can be written as

$$\ddot{x}_1 + 2\gamma_1 \dot{x}_1 + \omega_1^2 x_1 + \alpha_{22} x_2^2 = F\cos(\omega_D t) \tag{1}$$

$$\ddot{x}_2 + 2\gamma_2 \dot{x}_2 + \omega_2^2 x_2 + \alpha_{12} x_1 x_2 = 0 \tag{2}$$

Here, $\omega_1$ and $\omega_2$ are the resonance frequencies where $\omega_1 \approx 2\omega_2$, $\gamma_1$ and $\gamma_2$ are the damping rates, and $\alpha_{22} x_2^2$ and $\alpha_{12} x_1 x_2$ are the nonlinear coupling terms. This system is driven by $F\cos(\omega_D t)$, where $F$ is the drive amplitude and $\omega_D$ is the drive frequency. The Duffing, or third-order, nonlinearity was not included in eqs. (1) and (2) since experimental results showed the existence of phononic frequency combs before the Duffing nonlinearity had an effect on the frequency response [7]. Furthermore, this analysis is focused on existence conditions rather than accurate prediction of the comb amplitude. Solutions for $x_j$ are assumed to be $x_j = (u_j e^{i\omega_D t} + u_j^* e^{-i\omega_D t})/2$ for $j = 1, 2$, where $u_j$ are slowly varying envelopes [23,24]. After substituting these solutions into eqs (1) and (2), the equations of motion can then be written in terms of the complex amplitudes, $u_j$,

$$\ddot{u}_1 + [2\gamma_1 + 2i\omega_D]\dot{u}_1 + [(\omega_1^2 - \omega_D^2) + 2i\gamma_1\omega_D]u_1 + \alpha_{22}\frac{u_2^2}{2} = F \tag{3}$$

$$\ddot{u}_2 + [2\gamma_2 + i\omega_D]\dot{u}_2 + \left[\left(\omega_2^2 - \frac{\omega_D^2}{4}\right) + i\gamma_2\omega_D\right]u_2 + \frac{\alpha_{12}}{2}u_1 u_2^* = 0 \tag{4}$$

The generation of phononic frequency combs is indicated by the periodic modulation of the complex amplitudes $u_j$, where the modulation is slow compared to the carrier frequencies, resulting in pulsing between the two modes. In order to study these slow dynamics, the appropriate assumptions for the slowly varying envelope approximation are applied to eqs. (3) and (4) (see Supplementary Information for details on all derivations), providing two first-order differential equations that describe the normalized amplitudes of the two modes, $\psi_1$ and $\psi_2$, when driven by a single frequency near the resonance of mode 1 (i.e., $\omega_D \approx \omega_1$).

$$\frac{\partial \psi_1}{\partial \tau} = -if - (1 + i\Delta_1)\psi_1 + i\psi_2^2 \tag{5}$$

$$\frac{\partial \psi_2}{\partial \tau} = -(\gamma_{21} + i\Delta_2)\psi_2 + 2i\psi_1\psi_2^* \tag{6}$$

Here, $\tau = \gamma_1 t$, $f = \frac{\alpha_{12}}{8\gamma_1^2 \omega_D^2} F$, $\gamma_{21} = \frac{\gamma_2}{\gamma_1}$, $\Delta_1 = \frac{\omega_D - \omega_1}{\gamma_1}$, $\Delta_2 = \frac{\omega_D - 2\omega_2}{2\gamma_1}$, $\psi_1 = \frac{\alpha_{12}}{4\gamma_1 \omega_D} u_1$, and $\psi_2 = \frac{\sqrt{\alpha_{12}\alpha_{22}}}{4\gamma_2 \omega_D} u_2$. To understand the conditions for comb generation within these slow dynamics, the stationary points have been investigated. Assuming steady-state conditions (i.e., $\frac{\partial \psi_1}{\partial \tau} = \frac{\partial \psi_2}{\partial \tau} = 0$), eqs. (5) and (6) have two sets of stationary points: $(\psi_{1L}, \psi_{2L})$ and $(\psi_{1P}, \psi_{2P})$. The first set is $\psi_{1L} = -f/(-j + \Delta_1)$, $\psi_{2L} = 0$. These stationary points are stable when $f$ is small and provide the same expected amplitudes as the case where eqs. (1) and (2) are linear (i.e., $\alpha_{12} = \alpha_{22} = 0$). The second set of stationary points, $\psi_{1P}$ and $\psi_{2P}$, are defined by the following quadratic relationship,

$$|\psi_{2P}|^4 + (\gamma_{21} - \Delta_1\Delta_2)|\psi_{2P}|^2 + \frac{1}{4}(1 + \Delta_1^2)(\gamma_{21}^2 + \Delta_2^2) - f^2 = 0 \tag{7}$$

and $|\psi_{1P}|^2 = (\gamma_{21}^2 + \Delta_2^2)/4$. When $\psi_{1P}$ and $\psi_{2P}$ are stable stationary points, there is a 2:1 autoparametric resonance, or internal resonance, in which energy flows from mode 1 to mode 2, resulting in vibration of both modes with constant steady-state amplitude. Interestingly, $\psi_{1P}$ is not a function of $f$ because the amplitude is saturated and an increase in $f$ will only increase $\psi_{2P}$. Stability of the slow dynamics, eqs. (5) and (6), at $\psi_{1P}$ and $\psi_{2P}$ requires that the discriminant of eq. (7) be positive, $f \geq \frac{1}{2}|\gamma_{21}\Delta_1 + \Delta_2|$, which is shown in parameter space ($f$ vs. $\Delta_1$) in Fig. 2 (black line). This condition sets the boundary for autoparametric resonance, often referred to as an Arnold tongue, where $\psi_{1L}$ and $\psi_{2L}$ are stable stationary points below the line and $\psi_{1P}$ and $\psi_{2P}$ may be stable above the line. The transition from $\psi_{1L}$ and $\psi_{2L}$, to $\psi_{1P}$ and $\psi_{2P}$ is a Hopf bifurcation.

In the case of autoparametric resonance, the oscillatory amplitudes $\psi_1$ and $\psi_2$ are at steady state (i.e., the amplitudes remain constant over time). However, in the case of phononic frequency combs, we have previously shown that the amplitudes, $\psi_1$ and $\psi_2$, are modulated as a function of time such that periodic pulses are generated in the time domain, thereby resulting in frequency combs around the two modes [7]. This indicates that comb generation requires the slow dynamics to be unstable for some values of $\psi_{1P}$ and $\psi_{2P}$. Therefore eqs. (5) and (6) are linearized about these stationary points, such that small amplitude perturbations are defined as $\delta\psi_1$ and $\delta\psi_2$, and $\psi_1 \equiv \psi_{1P} + \delta\psi_1$ and $\psi_2 \equiv \psi_{2P} + \delta\psi_2$. Substituting into eqs. (5) and (6) and applying steady-state conditions, $\frac{\partial\psi_1}{\partial\tau} = \frac{\partial\psi_2}{\partial\tau} = 0$, the linearized dynamics can be written as follows.

$$\frac{\partial\delta\psi_1}{\partial\tau} = -(1 + i\Delta_1)\delta\psi_1 + 2i\psi_{2P}\delta\psi_2 \tag{8}$$

$$\frac{\partial\delta\psi_2}{\partial\tau} = -(\gamma_{21} + i\Delta_2)\delta\psi_2 + 2i\psi_{1P}\delta\psi_2^* + 2i\psi_{2P}^*\delta\psi_1 \tag{9}$$

In order to study the stability of the linearized dynamics, it is assumed that $\delta\psi_1 = b_1 e^{\lambda\gamma_1 t}$; $\delta\psi_1^* = b_2 e^{\lambda\gamma_1 t}$; $\delta\psi_2 = b_3 e^{\lambda\gamma_1 t}$ and $\delta\psi_2^* = b_4 e^{\lambda\gamma_1 t}$, and modulations $\delta\psi_1$ and $\delta\psi_2$ can only grow in strength if $\lambda$ is both real and positive. After applying the Routh-Hurwitz criterion [25] to analyze the stability of the linearized dynamics, eqs. (8) and (9), we obtain the following condition.

$$|\psi_{2P}|^2 \geq -\frac{\gamma_{21}(1 + \Delta_1^2)[1 + \Delta_1^2 + 4\gamma_{21}(1 + \gamma_{21})]}{4(1 + \gamma_{21})^2(1 + \Delta_1^2 + 2\gamma_{21} + 2\Delta_1\Delta_2)} \tag{10}$$

This condition dictates the minimum value of $|\psi_{2P}|^2$ that is required for non-zero values of $\delta\psi_1$ and $\delta\psi_2$. Only such non-zero amplitude modulations can ensure the generation of side-bands in the frequency domain, which in turn yields the frequency comb spectra. Hence, the energy exchange between modes 1 and 2 should be significant enough to enhance the value of $|\psi_{2P}|$ in order to generate frequency combs. Similar to the autoparametric resonance, the transition from stable amplitudes to amplitude modulation is also a Hopf bifurcation.

Since $|\psi_{2P}|^2$ is always positive (i.e., $|\psi_{2P}|$ is real), we obtain a boundary condition for the existence of phononic frequency combs as $2\Delta_1\Delta_2 \leq -(1 + \Delta_1^2 + 2\gamma_{21})$. This boundary forms the subset of the region of autoparametric resonance, $f \geq \frac{1}{2}|\gamma_{21}\Delta_1 + \Delta_2|$, as shown in Fig. 2, where the red line represents the threshold for instability of the linearized dynamics. While this analysis cannot prove that comb generation is the only dynamic behavior found within this instability bound, the frequency dependence of phononic frequency combs in experimental results [7] matches with the analytical evidence that the existence zone of phononic combs is bounded in drive frequency. Figure 2 shows that the phononic combs exist in the red-detuned and blue-detuned sides of driven mode 1 for $\omega_2 < \frac{\omega_1}{2}$ and $\omega_2 > \frac{\omega_1}{2}$, respectively, for the presented model.

In order to verify that this bounded region describes the conditions for phononic frequency combs, we conducted numerical simulations of eqs. (5) and (6) within this region. Figure 3 shows typical simulation results for mode amplitudes. The time domain responses (Figs. 3(a) and 3(b)) exhibit periodic oscillations and the corresponding fast Fourier transforms (FFT) (Figs. 3(c) and 3(d)) clearly demonstrate the existence of frequency combs.

Having established a description of the region in which combs can exist, we now investigate how the resonance frequencies and quality factors of modes 1 and 2 affect the location and shape of this region relative to the boundary describing the Hopf bifurcation to autoparametric resonance. To this end, the existence condition, eq. (10), is considered. From the resulting boundary line, the parameter $\omega_c = \frac{1}{2}(\omega_{D,min} + \omega_{D,max})$ is defined as the center frequency of the comb region (Fig. 2) (see Supplementary Information for definitions of $\omega_{D,max}$ and $\omega_{D,min}$).. It has been derived in the supplementary information, such that $\omega_c = \frac{1}{4}(3\omega_1 + 2\omega_2)$, with $\omega_2 \approx \frac{\omega_1}{2}$. For any value of $\omega_2$, the threshold for autoparametric resonance is always minimum when $\omega_D$ equals $\omega_1$. We now want to understand the minimum detuning from $\omega_1$ that is required to excite phononic frequency combs. We know that phononic combs only exist in a specific frequency band. Depending on whether $\omega_2 > \frac{\omega_1}{2}$ or $\omega_2 < \frac{\omega_1}{2}$, $\omega_1$ will be closer to either the left or right edge of phononic comb boundary. The difference between $\omega_1$ and the edge of the existence boundary for combs corresponds to the minimum detuning that is required for generating frequency combs, which is $\delta = |\omega_{D,edge} - \omega_1| = \frac{\omega_1}{\sqrt{2Q_1Q_2}}$. By increasing the quality factors, $Q_1$ and $Q_2$, $\delta$ is reduced, which in turn reduces the drive amplitude threshold for generating phononic combs. In other words, higher gain in the phononic combs can be obtained for smaller $\delta$. This analysis also shows that phononic combs can be generated only if the quality factor $Q_2$ is set above a critical value of $Q_{2,c} = \frac{2}{Q_1}\left(1 - \frac{2\omega_2}{\omega_1}\right)^{-2}$, as shown in Fig. 4a. The system parameters used in Fig. 4 were selected based on the experimental results in [7] so that the connection between the quality factors can be more easily understood. The frequency range $R$ corresponding to the existence band of phononic combs is found to increase with $Q_2$ as $R = \left|\omega_2 - \frac{\omega_1}{2}\right| - \left(\frac{\sqrt{2}\omega_1}{\sqrt{Q_1Q_2}}\right)$, which then asymptotes at $\left|\omega_2 - \frac{\omega_1}{2}\right|$ for large values of $Q_2$. Similar to $Q_2$, there also exists a critical value for $\left|\omega_2 - \frac{\omega_1}{2}\right|$, which is $g = 2\omega_1\sqrt{\frac{2}{Q_1 Q_2}}$. For $\left|\omega_2 - \frac{\omega_1}{2}\right| > g$, the frequency range $R$ scales linearly with $\left|\omega_2 - \frac{\omega_1}{2}\right|$, as shown in Fig. 4b. The above conditions can be used to design mechanical resonators that have sufficient quality factor and placement of resonance frequencies to systematically generate phononic frequency combs.

Equation (10) shows that there is only one boundary zone for phononic frequency combs in a two-mode system, which either lies on the red-detuned or blue-detuned side of mode 1 (i.e. either $\omega_D < \omega_1$ or $\omega_D > \omega_1$). There is an interesting discrepancy between this model and the experimental results shown in [7]. These results show that there are two boundary zones for phononic frequency combs and these zones lie on both sides of the resonance frequency (i.e., $\omega_D < \omega_1$ and $\omega_D > \omega_1$). The mode shapes for these two regions have been measured, as shown in [7], revealing that the mode coupling on either side of resonance is with two different modes. Referring to these as modes 2 and 3, the boundary zone that corresponds to $\omega_D < \omega_1$ can be explained by coupling between modes 1 and 2, and the zone corresponding to $\omega_D > \omega_1$ is due to coupling between modes 1 and 3. Hence, independently coupling a driven mode 1 to two different phonon modes leads to two bands of phononic frequency combs. Equation (10) can be directly employed to capture this more complex behavior, where the existence boundary for phononic combs resulting from the interactions of mode 1 and mode 2 is $2\Delta_1 \Delta_2 \leq -(1 + \Delta_1^2 + 2\gamma_{21})$ and between modes 1 and 3 is $2\Delta_1 \Delta_3 \leq -(1 + \Delta_1^2 + 2\gamma_{31})$.

In summary, this paper derives the existence conditions for phononic frequency comb generation with two coupled phonon modes in terms of drive frequency and amplitude. Using the boundary conditions, we investigated the influence of modal properties, including the quality factors and resonance frequencies of interacting modes on the conditions for comb generation. These include critical modal frequency separation, critical quality factors, and critical detuning that are required to produce a phononic frequency comb. For a system of two coupled phonon modes, the analysis revealed that there is only one existence zone for phononic combs. However, by correlating these analytical results with published experimental results, distinct existence boundaries of phononic frequency combs can be generated by independently coupling a driven mode with several other phonon modes. The results of this work will accelerate the development of mechanical devices with enhanced phononic comb properties for applications in sensing, communications and quantum information science.

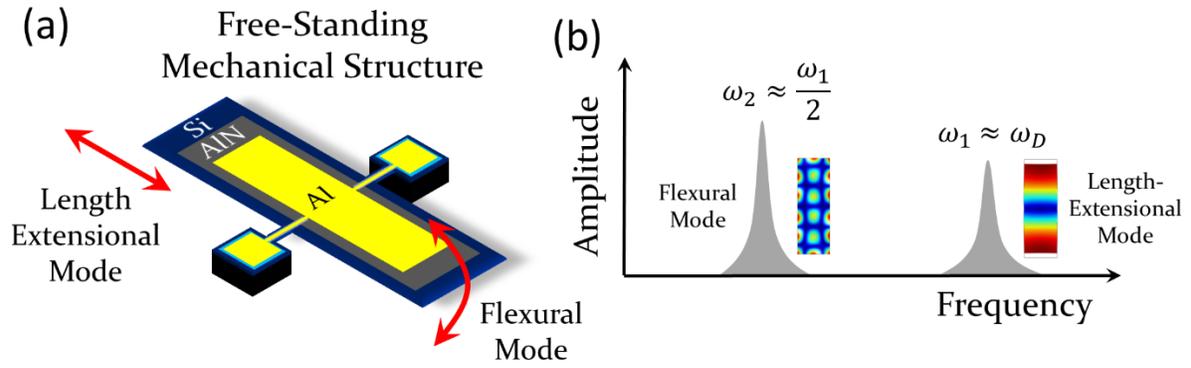

FIG. 1. Micromechanical resonator and phononic frequency comb concept. (a) Micromechanical resonator design used in [7] to generate phononic frequency combs. A length-extensional mode couples to a flexural mode through a nonlinear strain relationship. (b) Visual description of the mode coupling concept that results in phononic frequency combs, showing two modes where the resonance frequency of one mode is near double that of the other mode.

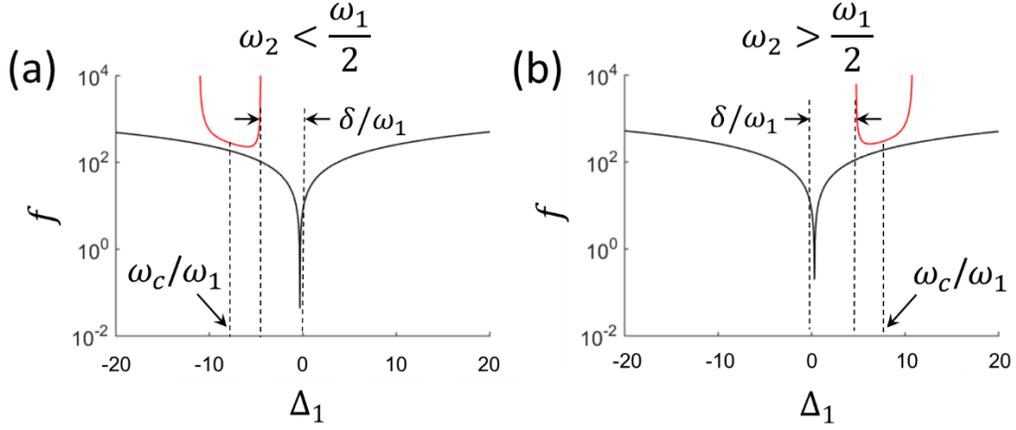

FIG. 2. Regions for parametric resonance (above the black line) and phononic frequency combs (above the red line) as a function of drive amplitude and frequency and the relative values of $\omega_1$ and $\omega_2$. That is, the drive conditions above the black line lead to parametric modal coupling and the drive conditions above the red line leads to phononic frequency combs. The existence bounds for both red-detuned resonances (a) and blue-detuned resonances (b) are shown, where the detuning is for $\omega_2$ relative to $\omega_1$. System parameters: (a) $\frac{\omega_1}{2\pi} = 3.86\ MHz$, $Q_1 = 1000$, $\frac{\omega_2}{2\pi} = 1.9\ MHz$ and $Q_2 = 10$; (b) $\frac{\omega_1}{2\pi} = 3.86\ MHz$, $Q_1 = 1000$, $\frac{\omega_2}{2\pi} = 1.96\ MHz$ and $Q_2 = 10$.

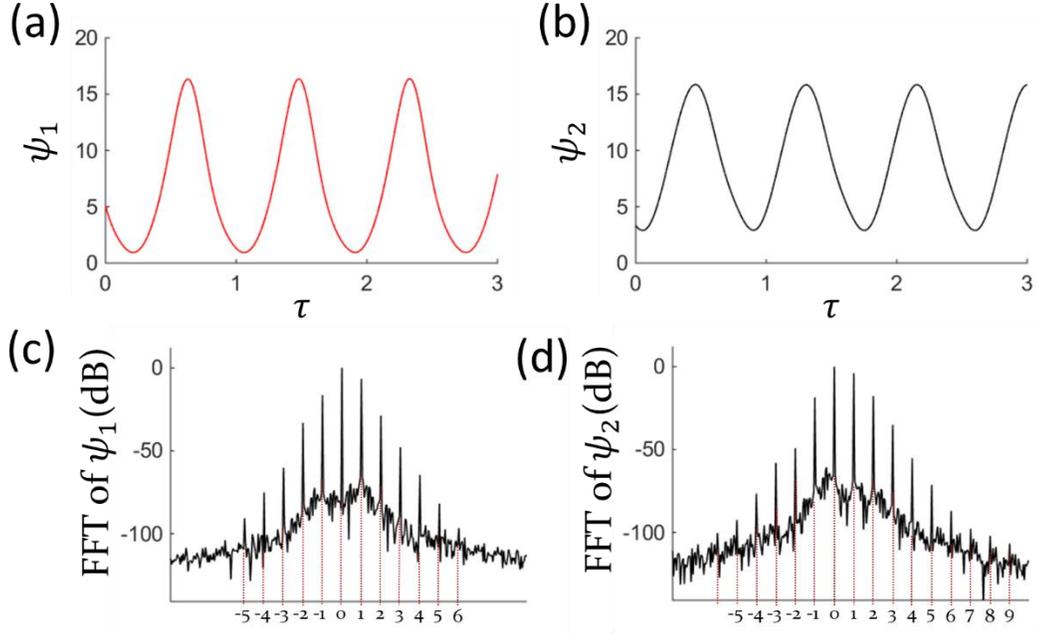

FIG. 3. Numerical simulation results for the mode amplitudes, eq. (9) and eq. (10), within the phononic frequency comb boundary. Simulation parameters: $\Delta_1 = 5$; $\kappa = -9$; $\Delta_2 = \frac{\Delta_1}{2} + \kappa$; $\gamma_{21} = 1$; $f = 20$. (a)-(b) Time domain responses. (c)-(d) Corresponding fast Fourier transforms (FFT) showing the existence of phononic frequency combs.

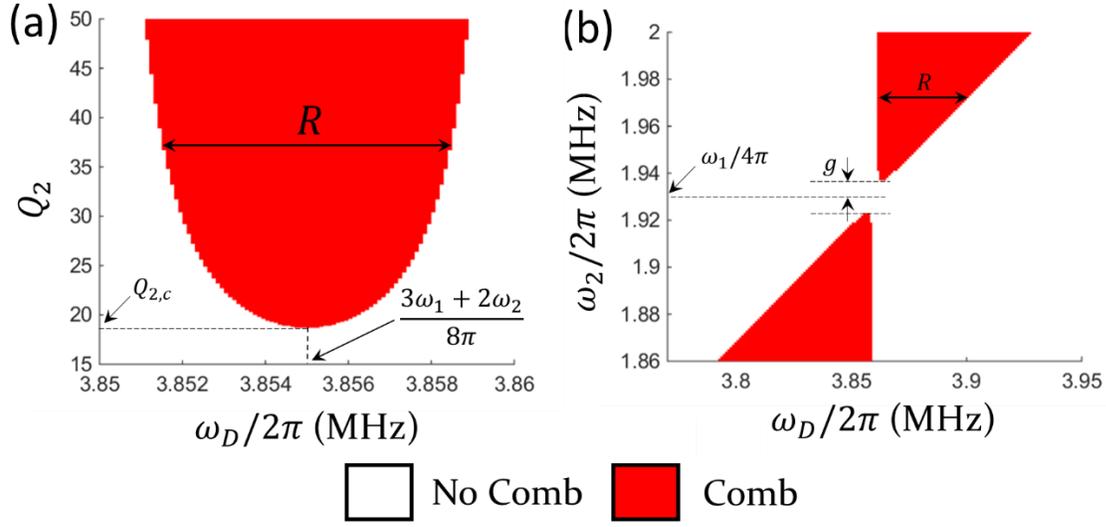

FIG. 4. Regions for generation of phononic frequency combs. (a) Quality factor of mode 2 vs. drive frequency. (b) Resonance frequency of mode 2 vs. drive frequency. System parameters: (a) $\frac{\omega_1}{2\pi} = 3.86\ MHz$, $Q_1 = 4000$; $\frac{\omega_2}{2\pi} = 1.92\ MHz$; (b) $\frac{\omega_1}{2\pi} = 3.86\ MHz$, $Q_1 = 4000$; $Q_2 = 50$.

# Supplementary Information

## Existence Conditions for Phononic Frequency Combs

Zhen Qi[1], Curtis R. Menyuk[1], Jason J. Gorman[2], and Adarsh Ganesan[2]

[1]Department of Computer Science and Electrical Engineering, University of Maryland, Baltimore County, Baltimore, MD 21250, USA

[2]National Institute of Standards and Technology, Gaithersburg, MD 20899, USA

## S1. Derivation of Boundary Line for Parametric Resonance and Phononic Frequency Combs

The equations of motion corresponding to two coupled modes are given by

$$\ddot{x}_1 + 2\gamma_1 \dot{x}_1 + \omega_1^2 x_1 + \alpha_{22} x_2^2 = F\cos(\omega_D t) \tag{1.1}$$
$$\ddot{x}_2 + 2\gamma_2 \dot{x}_2 + \omega_2^2 x_2 + \alpha_{12} x_1 x_2 = 0 \tag{1.2}$$

The modes are written as

$$x_1 = \frac{1}{2}\left(u_1 e^{i\omega_D t} + u_1^* e^{-i\omega_D t}\right) \tag{1.3}$$
$$x_2 = \frac{1}{2}\left(u_2 e^{i\frac{\omega_D}{2} t} + u_2^* e^{-i\frac{\omega_D}{2} t}\right) \tag{1.4}$$

The derivatives $\dot{x}_1$, $\dot{x}_2$, $\ddot{x}_1$ and $\ddot{x}_2$ and the coupling terms $x_1 x_2$ and $x_2^2$ are obtained as

$$\dot{x}_1 = \frac{1}{2}\left(i\omega_D u_1 e^{i\omega_D t} + \dot{u}_1 e^{i\omega_D t} - i\omega_D u_1 e^{-i\omega_D t} + \dot{u}_1^* e^{-i\omega_D t}\right) \tag{1.5}$$

$$\dot{x}_2 = \frac{1}{2}\left(i\frac{\omega_D}{2} u_2 e^{i\frac{\omega_D}{2} t} + \dot{u}_2 e^{i\frac{\omega_D}{2} t} - i\frac{\omega_D}{2} u_2 e^{-i\frac{\omega_D}{2} t} + \dot{u}_2^* e^{-i\frac{\omega_D}{2} t}\right) \tag{1.6}$$

$$\ddot{x}_1 = \frac{1}{2}\left(\ddot{u}_1 e^{i\omega_D t} + 2i\omega_D \dot{u}_1 e^{i\omega_D t} - \omega_D^2 u_1 e^{i\omega_D t} + \ddot{u}_1^* e^{-i\omega_D t} - 2i\omega_D \dot{u}_1^* e^{-i\omega_D t} - \omega_D^2 u_1^* e^{-i\omega_D t}\right) \tag{1.7}$$

$$\ddot{x}_2 = \frac{1}{2}\left(\ddot{u}_2 e^{i\frac{\omega_D}{2} t} + i\omega_D \dot{u}_2 e^{i\frac{\omega_D}{2} t} - \frac{\omega_D^2}{4} u_2 e^{i\frac{\omega_D}{2} t} + \ddot{u}_2^* e^{-i\frac{\omega_D}{2} t} - i\omega_D \dot{u}_2^* e^{-i\frac{\omega_D}{2} t} - \frac{\omega_D^2}{4} u_2^* e^{-i\frac{\omega_D}{2} t}\right) \tag{1.8}$$

$$x_1 x_2 = \frac{1}{4}\left(u_1^* u_2^* e^{-3i\frac{\omega_D}{2} t} + u_1^* u_2 e^{-i\frac{\omega_D}{2} t} + u_1 u_2^* e^{i\frac{\omega_D}{2} t} + u_1 u_2 e^{3i\frac{\omega_D}{2} t}\right) \tag{1.9}$$

$$x_2^2 = \frac{1}{4}\left(u_2^2 e^{i\omega_D t} + u_2^{*2} e^{-i\omega_D t} + 2|u_2|^2\right) \tag{1.10}$$

Now, by taking $\cos(\omega_D t) = \frac{1}{2}\left(e^{i\omega_D t} + e^{-i\omega_D t}\right)$, eq. (1.1) and eq. (1.2) become

$$\begin{aligned}&\frac{1}{2}\left(\ddot{u}_1 e^{i\omega_D t} + 2i\omega_D \dot{u}_1 e^{i\omega_D t} - \omega_D^2 u_1 e^{i\omega_D t} + \ddot{u}_1^* e^{-i\omega_D t} - 2i\omega_D \dot{u}_1^* e^{-i\omega_D t} - \omega_D^2 u_1^* e^{-i\omega_D t}\right) \\ &+ \gamma_1\left(i\omega_D u_1 e^{i\omega_D t} + \dot{u}_1 e^{i\omega_D t} - i\omega_D u_1 e^{-i\omega_D t} + \dot{u}_1^* e^{-i\omega_D t}\right) \\ &+ \frac{\omega_1^2}{2}\left(u_1 e^{i\omega_D t} + u_1^* e^{-i\omega_D t}\right) + \frac{\alpha_{22}}{4}\left(u_2^2 e^{i\omega_D t} + u_2^{*2} e^{-i\omega_D t} + 2|u_2|^2\right) \\ &= \frac{F}{2}\left(e^{i\omega_D t} + e^{-i\omega_D t}\right)\end{aligned} \tag{1.11}$$

$$\frac{1}{2}\left(\ddot{u}_2 e^{i\frac{\omega_D}{2}t} + i\omega_D \dot{u}_2 e^{i\frac{\omega_D}{2}t} - \frac{\omega_D^2}{4} u_2 e^{i\frac{\omega_D}{2}t} + \ddot{u}_2^* e^{-i\frac{\omega_D}{2}t} - i\omega_D \dot{u}_2^* e^{-i\frac{\omega_D}{2}t} - \frac{\omega_D^2}{4} u_2^* e^{-i\frac{\omega_D}{2}t}\right)$$
$$+ \gamma_2 \left(i\frac{\omega_D}{2} u_2 e^{i\frac{\omega_D}{2}t} + \dot{u}_2 e^{i\frac{\omega_D}{2}t} - i\frac{\omega_D}{2} u_2 e^{-i\frac{\omega_D}{2}t} + \dot{u}_2^* e^{-i\frac{\omega_D}{2}t}\right)$$
$$+ \frac{\omega_2^2}{2}\left(u_2 e^{i\frac{\omega_D}{2}t} + u_2^* e^{-i\frac{\omega_D}{2}t}\right)$$
$$+ \frac{\alpha_{12}}{4}\left(u_1^* u_2^* e^{-3i\frac{\omega_D}{2}t} + u_1^* u_2 e^{-i\frac{\omega_D}{2}t} + u_1 u_2^* e^{i\frac{\omega_D}{2}t} + u_1 u_2 e^{3i\frac{\omega_D}{2}t}\right) = 0 \quad (1.12)$$

We now collect the coefficients of $e^{i\omega_D t}$ in eq. (1.11) and coefficients of $e^{i\frac{\omega_D}{2}t}$ in eq. (1.12) to obtain the following relationships.

$$\ddot{u}_1 + [2\gamma_1 + 2i\omega_D]\dot{u}_1 + [(\omega_1^2 - \omega_D^2) + 2i\gamma_1\omega_D]u_1 + \alpha_{22}\frac{u_2^2}{2} = F \quad (1.13)$$

$$\ddot{u}_2 + [2\gamma_2 + i\omega_D]\dot{u}_2 + \left[\left(\omega_2^2 - \frac{\omega_D^2}{4}\right) + i\gamma_2\omega_D\right]u_2 + \frac{\alpha_{12}}{2}u_1 u_2^* = 0 \quad (1.14)$$

We now multiply eq. (1.13) by $\frac{\alpha_{12}}{2\omega_D^4}$ and eq. (1.14) by $\frac{\sqrt{\alpha_{12}\alpha_{22}}}{2\omega_D^4}$.

$$\frac{\partial^2\left(\frac{\alpha_{12}}{2\omega_D^2}u_1\right)}{\partial(\omega_D t)^2} + \left[2\left(\frac{\gamma_1}{\omega_D}\right) + 2i\right]\dot{u}_1 + \left[\left(\frac{\omega_1}{\omega_D}\right)^2 - 1 + 2i\left(\frac{\gamma_1}{\omega_D}\right)\right]\frac{\partial\left(\frac{\alpha_{12}}{2\omega_D^2}u_1\right)}{\partial(\omega_D t)}$$
$$+ \left(\frac{\sqrt{\alpha_{12}\alpha_{22}}}{2\omega_D^2}u_2\right)^2 = \left(\frac{\alpha_{12}}{2\omega_D^4}F\right) \quad (1.15)$$

$$\frac{\partial^2\left(\frac{\sqrt{\alpha_{12}\alpha_{22}}}{2\omega_D^2}u_2\right)}{\partial(\omega_D t)^2} + \left[2\left(\frac{\gamma_2}{\omega_D}\right) + i\right]\dot{u}_2 + \left[\left(\frac{\omega_2}{\omega_D}\right)^2 - \frac{1}{4} + i\left(\frac{\gamma_2}{\omega_D}\right)\right]\frac{\partial\left(\frac{\sqrt{\alpha_{12}\alpha_{22}}}{2\omega_D^2}u_2\right)}{\partial(\omega_D t)}$$
$$+ \left(\frac{\alpha_{12}}{2\omega_D^2}u_1\right)\left(\frac{\sqrt{\alpha_{12}\alpha_{22}}}{2\omega_D^2}u_2^*\right) = 0 \quad (1.16)$$

Now, we apply the following transformation and hence obtain eq. (1.18) and eq. (1.19).

$$T = \omega_D t, \quad \bar{F} = \frac{\alpha_{12}}{2\omega_D^4}F$$
$$\bar{\psi}_1 = \frac{\alpha_{12}}{2\omega_D^2}u_1, \quad \bar{\psi}_2 = \frac{\sqrt{\alpha_{12}\alpha_{22}}}{2\omega_D^2}u_2 \quad (1.17)$$
$$\bar{\omega}_1 = \frac{\omega_1}{\omega_D}, \quad \bar{\omega}_2 = \frac{\omega_2}{\omega_D}$$
$$\bar{\gamma}_1 = \frac{\gamma_1}{\omega_D}, \quad \bar{\gamma}_2 = \frac{\gamma_2}{\omega_D}$$

$$\frac{\partial^2 \bar{\psi}_1}{\partial T^2} + [2\bar{\gamma}_1 + 2i]\frac{\partial \bar{\psi}_1}{\partial T} + [\bar{\omega}_1^2 - 1 + 2i\bar{\gamma}_1]\bar{\psi}_1 + \bar{\psi}_2^2 = \bar{F} \quad (1.18)$$

$$\frac{\partial^2 \bar{\psi}_2}{\partial T^2} + [2\bar{\gamma}_2 + i]\frac{\partial \bar{\psi}_2}{\partial T} + \left[\bar{\omega}_2^2 - \frac{1}{4} + i\bar{\gamma}_2\right]\bar{\psi}_2 + \bar{\psi}_1\bar{\psi}_2^* = 0 \quad (1.19)$$

For slowly varying fields, we assume the following conditions and hence we obtain eq. (1.23) and eq. (1.24).

$$\left(\frac{\partial^2}{\partial T^2}\right) \to 0 \tag{1.20}$$

$$\left(\bar{\gamma}_1 \frac{\partial}{\partial T}\right) \to 0, \left(\bar{\gamma}_2 \frac{\partial}{\partial T}\right) \to 0 \tag{1.21}$$

$$\bar{\omega}_1 \to 1, \bar{\omega}_2 \to \frac{1}{2} \tag{1.22}$$

$$2i\frac{\partial \bar{\psi}_1}{\partial T} + 2i\bar{\gamma}_1 \bar{\psi}_1 + \bar{\psi}_2^2 = \bar{F} \tag{1.23}$$

$$i\frac{\partial \bar{\psi}_2}{\partial T} + i\bar{\gamma}_2 \bar{\psi}_2 + \bar{\psi}_1 \bar{\psi}_2^* = 0 \tag{1.24}$$

We now multiply eq. (1.23) by $\frac{1}{4\bar{\gamma}_1^2}$ and eq. (1.24) by $\frac{1}{2\bar{\gamma}_1^2}$.

$$\frac{\partial \left(\frac{\bar{\psi}_1}{2\bar{\gamma}_1}\right)}{\partial (\bar{\gamma}_1 T)} = -\left(\frac{\bar{\psi}_1}{2\bar{\gamma}_1}\right) + i\left(\frac{\bar{\psi}_2}{2\bar{\gamma}_1}\right)^2 - i\left(\frac{\bar{F}}{4\bar{\gamma}_1^2}\right) \tag{1.25}$$

$$\frac{\partial \left(\frac{\bar{\psi}_2}{2\bar{\gamma}_1}\right)}{\partial (\bar{\gamma}_1 T)} = -\left(\frac{\bar{\gamma}_2}{\bar{\gamma}_1}\right)\left(\frac{\bar{\psi}_2}{2\bar{\gamma}_1}\right) + 2i\left(\frac{\bar{\psi}_1}{2\bar{\gamma}_1}\right)\left(\frac{\bar{\psi}_2^*}{2\bar{\gamma}_1}\right) \tag{1.26}$$

Now, we make the following transformations.

$$\tau = \bar{\gamma}_1 T; f = \frac{\bar{F}}{4\bar{\gamma}_1^2}; \gamma_{21} = \frac{\bar{\gamma}_2}{\bar{\gamma}_1}; \kappa = \frac{\bar{\omega}_1 - 2\bar{\omega}_2}{2\bar{\gamma}_1} \tag{1.27}$$

$$\Delta_1 = \frac{1-\bar{\omega}_1}{\bar{\gamma}_1}; \Delta_2 = \frac{\frac{1}{2}-\bar{\omega}_2}{\bar{\gamma}_1} = \frac{\Delta_1}{2} + \kappa \tag{1.28}$$

$$\frac{\bar{\psi}_1}{2\bar{\gamma}_1} = \tilde{\psi}_1 = \psi_1 e^{i\Delta_1 \tau}; \frac{\bar{\psi}_2}{2\bar{\gamma}_2} = \tilde{\psi}_2 = \psi_2 e^{i\Delta_2 \tau} \tag{1.29}$$

$$\frac{\partial \tilde{\psi}_1}{\partial \tau} = \frac{\partial \psi_1}{\partial \tau} e^{i\Delta_1 \tau} + i\Delta_1 \psi_1 \tag{1.30}$$

$$\frac{\partial \tilde{\psi}_2}{\partial \tau} = \frac{\partial \psi_2}{\partial \tau} e^{i\Delta_2 \tau} + i\Delta_2 \psi_2 \tag{1.31}$$

In the limiting cases of $\Delta_1 \to 0$ and $\Delta_2 \to 0$, we obtain the following.

$$\frac{\bar{\psi}_1}{2\bar{\gamma}_1} = \tilde{\psi}_1 = \psi_1; \frac{\bar{\psi}_2}{2\bar{\gamma}_2} = \tilde{\psi}_2 = \psi_2 \tag{1.32}$$

$$\frac{\partial \tilde{\psi}_1}{\partial \tau} = \frac{\partial \psi_1}{\partial \tau} + i\Delta_1 \psi_1 \tag{1.33}$$

$$\frac{\partial \tilde{\psi}_2}{\partial \tau} = \frac{\partial \psi_2}{\partial \tau} + i\Delta_2 \psi_2 \tag{1.34}$$

Hence, we re-write eq. (1.25) and eq. (1.26) as

$$\frac{\partial \psi_1}{\partial \tau} = -if - (1 + i\Delta_1)\psi_1 + i\psi_2^2 \tag{1.35}$$

$$\frac{\partial \psi_2}{\partial \tau} = -(\gamma_{21} + i\Delta_2)\psi_2 + 2i\psi_1\psi_2^* \tag{1.36}$$

The stationary values are obtained by setting $\frac{\partial \psi_1}{\partial \tau}$ and $\frac{\partial \psi_2}{\partial \tau}$ to zero.

$$-if - (1 + i\Delta_1)\psi_1 + i\psi_2^2 = 0 \tag{1.37}$$
$$-(\gamma_{21} + i\Delta_2)\psi_2 + 2i\psi_1\psi_2^* = 0 \tag{1.38}$$

The eq. (1.37) and eq. (1.38) possess a trivial solution of eq. (1.39), and a non-trivial solution of eq. (1.40).

$$\psi_{1L} = \frac{-f}{-i + \Delta_1} \tag{1.39a}$$

$$\psi_{2L} = 0 \tag{1.39b}$$

$$\psi_{1P} = i\frac{-f + \psi_{2P}^2}{(1 + i\Delta_1)} \tag{1.40}$$

By substituting eq. (1.40) in eq. (1.38), we get

$$
\begin{aligned}
&-(\gamma_{21} + i\Delta_2)\psi_{2P} - 2\frac{-f + \psi_{2P}^2}{(1 + i\Delta_1)}\psi_{2P}^* = 0 \\
&\Rightarrow -(\gamma_{21} + i\Delta_2)(1 + i\Delta_1)\psi_{2P} + 2f\psi_{2P}^* - 2|\psi_{2P}|^2\psi_{2P} = 0 \\
&\Rightarrow [(\gamma_{21} + i\Delta_2)(1 + i\Delta_1) + 2|\psi_{2P}|^2]\psi_{2P} = 2f\psi_{2P}^* \\
&\Rightarrow \psi_{2P} = \frac{2f}{[(\gamma_{21} + i\Delta_2)(1 + i\Delta_1) + 2|\psi_{2P}|^2]}\psi_{2P}^* \\
&\Rightarrow |\psi_{2P}|^2 = \frac{4f^2}{[(\gamma_{21} + i\Delta_2)(1 + i\Delta_1) + 2|\psi_{2P}|^2][(\gamma_{21} - i\Delta_2)(1 - i\Delta_1) + 2|\psi_{2P}|^2]}|\psi_{2P}|^2 \\
&\Rightarrow \frac{1}{4}[(\gamma_{21} + i\Delta_2)(1 + i\Delta_1) + 2|\psi_{2P}|^2][(\gamma_{21} - i\Delta_2)(1 - i\Delta_1) + 2|\psi_{2P}|^2] = f^2 \\
&\Rightarrow |\psi_{2P}|^4 + (\gamma_{21} - \Delta_1\Delta_2)|\psi_{2P}|^2 + \frac{1}{4}(1 + \Delta_1^2)(\gamma_{21}^2 + \Delta_2^2) - f^2 = 0
\end{aligned}
\tag{1.41}
$$

The solution for $\psi_{1P}$ can be obtained by rewriting eq. (1.41) as

$$
\begin{aligned}
-(\gamma_{21} + i\Delta_2)\psi_{2P} + 2i\psi_{1P}\psi_{2P}^* &= 0 \\
2i\psi_{1P}\psi_{2P}^* &= (\gamma_{21} + i\Delta_2)\psi_{2P} \\
2i\psi_{1P}^*\psi_{2P} &= (\gamma_{21} - i\Delta_2)\psi_{2P}^* \\
4|\psi_{1P}|^2 &= (\gamma_{21}^2 + \Delta_2^2)
\end{aligned}
\tag{1.42}
$$

For real solutions of $|\psi_{2P}|^2$, the discriminant must be non-negative. Hence, we obtain the following condition.

$$(\gamma_{21} - \Delta_1\Delta_2)^2 - (1 + \Delta_1^2)(\gamma_{21}^2 + \Delta_2^2) + 4f^2 \geq 0$$
$$\Rightarrow -2\gamma_{21}\Delta_1\Delta_2 - \gamma_{21}^2\Delta_1^2 - \Delta_2^2 + 4f^2 \geq 0$$
$$\Rightarrow f^2 \geq \frac{1}{4}(\gamma_{21}\Delta_1 + \Delta_2)^2 \tag{1.43}$$
$$\Rightarrow f \geq \frac{1}{2}|\gamma_{21}\Delta_1 + \Delta_2|$$

The minimum value of $|\psi_{2P}|^2$ should also be non-negative, and hence we obtain another condition.

$$-\frac{(\gamma_{21} - \Delta_1\Delta_2)}{2} \geq 0$$
$$\Delta_1\Delta_2 \geq \gamma_{21} \tag{1.44}$$

Now, by taking fluctuations around the stationary solutions of eq. (1.41) and eq. (1.42), we redefine the fields as

$$\psi_1 = \psi_{1P} + \delta\psi_1 \tag{1.45a}$$
$$\psi_2 = \psi_{2P} + \delta\psi_2 \tag{1.45b}$$

Here, the stationary solutions $\psi_{1P}$ and $\psi_{2P}$ satisfy eq. (1.37) and eq. (1.38) and therefore we have the following.

$$-if - (1 + i\Delta_1)\psi_{1P} + i\psi_{2P}^2 = 0 \tag{1.46}$$
$$-(\gamma_{21} + i\Delta_2)\psi_{2P} + 2i\psi_{1P}\psi_{2P}^* = 0 \tag{1.47}$$

By substituting eq. (1.45), eq. (1.46) and eq. (1.47), we re-write eq. (1.35) and eq. (1.36) as

$$\frac{\partial \delta\psi_1}{\partial \tau} = -(1 + i\Delta_1)\delta\psi_1 + 2i\psi_{2P}\delta\psi_2 + i(\delta\psi_2)^2 \tag{1.48}$$

We ignore second-order fluctuations and rewrite eq. (1.48) as

$$\frac{\partial \delta\psi_1}{\partial \tau} = -(1 + i\Delta_1)\delta\psi_1 + 2i\psi_{2P}\delta\psi_2 \tag{1.49}$$

Similarly, we also obtain a dynamical equation for $\delta\psi_2$.

$$\frac{\partial \delta\psi_2}{\partial \tau} = -(\gamma_{21} + i\Delta_2)\delta\psi_2 + 2i\psi_{1P}\delta\psi_2^* + 2i\psi_{2P}^*\delta\psi_1 \tag{1.50}$$

The dynamical equations for $\delta\psi_1^*$ and $\delta\psi_2^*$ can also be written as

$$\frac{\partial \delta\psi_1^*}{\partial \tau} = -(1 - i\Delta_1)\delta\psi_1^* - 2i\psi_{2P}^*\delta\psi_2^* \tag{1.51}$$

$$\frac{\partial \delta\psi_2^*}{\partial \tau} = -(\gamma_{21} - i\Delta_2)\delta\psi_2^* - 2i\psi_{1P}^*\delta\psi_2 + 2i\psi_{2P}\delta\psi_1^* \tag{1.52}$$

We can now make the following assumptions to eq. (1.49), eq. (1.50), eq. (1.51) and eq. (1.52) and hence obtain eq. (1.57), eq. (1.58), eq. (1.59) and eq. (1.60).

$$\delta\psi_1 = b_1 e^{\lambda\tau} \tag{1.53}$$
$$\delta\psi_1^* = b_2 e^{\lambda\tau} \tag{1.54}$$
$$\delta\psi_2 = b_3 e^{\lambda\tau} \tag{1.55}$$
$$\delta\psi_2^* = b_4 e^{\lambda\tau} \tag{1.56}$$

$$\lambda b_1 e^{\lambda t} = -(1 + i\Delta_1)b_1 e^{\lambda\tau} + 2i\psi_{2P} b_3 e^{\lambda\tau}$$
$$\Rightarrow b_1 = \frac{2i\psi_{2P}}{\lambda + (1 + i\Delta_1)} b_3 \tag{1.57}$$

$$\lambda b_2 e^{\lambda t} = -(1 - i\Delta_1)b_2 e^{\lambda\tau} - 2i\psi_{2P}^* b_4 e^{\lambda\tau}$$
$$\Rightarrow b_2 = \frac{-2i\psi_{2P}^*}{\lambda + (1 - i\Delta_1)} b_4 \tag{1.58}$$

$$\lambda b_3 e^{\lambda t} = -(\gamma_{21} + i\Delta_2)b_3 e^{\lambda t} + 2i\psi_{1P} b_4 e^{\lambda t} + 2i\psi_{2P}^* b_1 e^{\lambda t}$$
$$\Rightarrow \lambda b_3 + (\gamma_{21} + i\Delta_2)b_3 - 2i\psi_{2P}^* b_1 = 2i\psi_{1P} b_4$$
$$\Rightarrow \lambda b_3 + (\gamma_{21} + i\Delta_2)b_3 - 2i\psi_{2P}^* \frac{2i\psi_{2P}}{\lambda + (1 + i\Delta_1)} b_3 = 2i\psi_{1P} b_4$$
$$\Rightarrow \frac{\{\lambda + (\gamma_{21} + i\Delta_2)\}\{\lambda + (1 + i\Delta_1)\} + 4|\psi_{2P}|^2}{\lambda + (1 + i\Delta_1)} b_3 = 2i\psi_{1P} b_4 \tag{1.59}$$
$$\Rightarrow b_3 = \frac{2i\psi_{1P}\{\lambda + (1 + i\Delta_1)\}}{\{\lambda + (\gamma_{21} + i\Delta_2)\}\{\lambda + (1 + i\Delta_1)\} + 4|\psi_{2P}|^2} b_4$$

$$\lambda b_4 e^{\lambda t} = -(\gamma_{21} - i\Delta_2)b_4 e^{\lambda t} - 2i\psi_{2P} b_2 e^{\lambda t} - 2i\psi_{1P}^* b_3 e^{\lambda t}$$
$$\Rightarrow \lambda b_4 + (\gamma_{21} - i\Delta_2)b_4 + 2i\psi_{2P} \frac{-2i\psi_{2P}^*}{\lambda + (1 - i\Delta_1)} b_4$$
$$+ 2i\psi_{1P}^* \frac{2i\bar{\psi}_{1P}\{\lambda + (1 + i\Delta_1)\}}{\{\lambda + (\gamma_{21} + i\Delta_2)\}\{\lambda + (1 + i\Delta_1)\} + 4|\psi_{2P}|^2} b_4 = 0 \tag{1.60}$$
$$\Rightarrow \frac{\{\lambda + (\gamma_{21} - i\Delta_2)\}\{\lambda + (1 - i\Delta_1)\} + 4|\psi_{2P}|^2}{\lambda + (1 - i\Delta_1)}$$
$$= \frac{4|\psi_{1P}|^2\{\lambda + (1 + i\Delta_1)\}}{\{\lambda + (\gamma_{21} + i\Delta_2)\}\{\lambda + (1 + i\Delta_1)\} + 4|\psi_{2P}|^2}$$

$$[\{\lambda + (\gamma_{21} - i\Delta_2)\}\{\lambda + (1 - i\Delta_1)\} + 4|\psi_{2P}|^2][\{\lambda + (\gamma_{21} + i\Delta_2)\}\{\lambda + (1 + i\Delta_1)\} + 4|\psi_{2P}|^2]$$
$$= 4|\psi_{1P}|^2\{\lambda + (1 + i\Delta_1)\}\{\lambda + (1 - i\Delta_1)\} \tag{1.61}$$

$$\Rightarrow \{\lambda^2 + [(\gamma_{21} - i\Delta_2) + (1 - i\Delta_1)]\lambda + (\gamma_{21} - i\Delta_2)(1 - i\Delta_1) + 4|\psi_{2P}|^2\}\{\lambda^2$$
$$+ [(\gamma_{21} + i\Delta_2) + (1 + i\Delta_1)]\lambda + (\gamma_{21} + i\Delta_2)(1 + i\Delta_1) + 4|\psi_{2P}|^2\}$$
$$= 4|\psi_{1P}|^2\{\lambda^2 + 2\lambda + 1 + \Delta_1^2\}$$

$$\Rightarrow \left\{ \begin{array}{c} \lambda^4 \\ +[(\gamma_{21} - i\Delta_2) + (1 - i\Delta_1) + (\gamma_{21} + i\Delta_2) + (1 + i\Delta_1)]\lambda^3 \\ +\left[ \begin{array}{c} [(\gamma_{21} - i\Delta_2) + (1 - i\Delta_1)][(\gamma_{21} + i\Delta_2) + (1 + i\Delta_1)] \\ +(\gamma_{21} - i\Delta_2)(1 - i\Delta_1) + (\gamma_{21} + i\Delta_2)(1 + i\Delta_1) + 8|\psi_{2P}|^2 - 4|\psi_{1P}|^2 \end{array} \right]\lambda^2 \\ +[4[(\gamma_{21} - i\Delta_2) + (1 - i\Delta_1) + (\gamma_{21} + i\Delta_2) + (1 + i\Delta_1)]|\psi_{2P}|^2 - 8|\psi_{1P}|^2]\lambda \\ +[16|\psi_{2P}|^4 + 8|\psi_{2P}|^2[\gamma_{21} - \Delta_1\Delta_2] + (1 + \Delta_1^2)(\gamma_{21}^2 + \Delta_2^2 - 4|\psi_{1P}|^2)] \end{array} \right\} = 0$$

$$\Rightarrow \left\{ \begin{array}{c} \lambda^4 \\ +[2(1 + \gamma_{21})]\lambda^3 \\ +[1 + \Delta_1^2 + \gamma_{21}^2 + \Delta_2^2 + 4\gamma_{21} + 8|\psi_{2P}|^2 - 4|\psi_{1P}|^2]\lambda^2 \\ +[8(1 + \gamma_{21})|\psi_{2P}|^2 + 2(\gamma_{21}^2 + \Delta_2^2) + (2\gamma_{21})(1 + \Delta_1^2) - 8|\psi_{1P}|^2]\lambda \\ +[16|\psi_{2P}|^4 + 8|\psi_{2P}|^2(\gamma_{21} - \Delta_1\Delta_2) + (1 + \Delta_1^2)(\gamma_{21}^2 + \Delta_2^2 - 4|\psi_{1P}|^2)] \end{array} \right\} = 0 \quad (1.62)$$

By substituting eq. (1.42) in eq. (1.62), we get

$$\Rightarrow \left\{ \begin{array}{c} \lambda^4 \\ +[2(1 + \gamma_{21})]\lambda^3 \\ +[1 + \Delta_1^2 + 4\gamma_{21} + 8|\psi_{2P}|^2]\lambda^2 \\ +[8(1 + \gamma_{21})|\psi_{2P}|^2 + (2\gamma_{21})(1 + \Delta_1^2)]\lambda \\ +[16|\psi_{2P}|^4 + 8|\psi_{2P}|^2(\gamma_{21} - \Delta_1\Delta_2)] \end{array} \right\} = 0 \quad (1.63)$$

Eq. (1.63) can be written as

$$\lambda^4 + c_1\lambda^3 + c_2\lambda^2 + c_3\lambda + c_4 = 0 \quad (1.64)$$

where

$$\begin{array}{c} c_1 = 2(1 + \gamma_{21}) \\ c_2 = 1 + \Delta_1^2 + 4\gamma_{21} + 8|\psi_{2P}|^2 \\ c_3 = 8(1 + \gamma_{21})|\psi_{2P}|^2 + (2\gamma_{21})(1 + \Delta_1^2) \\ c_4 = 16|\psi_{2P}|^4 + 8|\psi_{2P}|^2(\gamma_{21} - \Delta_1\Delta_2) \end{array} \quad (1.65)$$

According to Routh-Hurwitz criterion,

$$c_1 \geq 0, c_2 \geq 0, c_3 \geq 0, c_4 \geq 0,$$
$$c_1c_2 - c_3 \geq 0 \quad (1.66)$$
$$c_3(c_1c_2 - c_3) - c_1^2 c_4 \leq 0$$

Here, $c_1 \geq 0, c_2 \geq 0, c_3 \geq 0, c_1 c_2 - c_3 \geq 0$ is always satisfied. The condition $c_4 \geq 0$ is satisfied when $|\psi_{2P}|^2$ is always positive. We now simplify the expression $c_3(c_1 c_2 - c_3) - c_1^2 c_4 \leq 0$ to obtain a condition for $|\psi_{2P}|^2$.

$$c_3(c_1 c_2 - c_3) - c_1^2 c_4 \leq 0$$

$$\Rightarrow [8(1+\gamma_{21})|\psi_{2P}|^2 + 2\gamma_{21}(1+\Delta_1^2)]\Big[2(1+\gamma_{21})(1+\Delta_1^2 + 4\gamma_{21} + 8|\psi_{2P}|^2)$$
$$- [8(1+\gamma_{21})|\psi_{2P}|^2 + 2\gamma_{21}(1+\Delta_1^2)]\Big]$$
$$- 4(1+\gamma_{21})^2[16|\psi_{2P}|^4 + 8|\psi_{2P}|^2(\gamma_{21} - \Delta_1\Delta_2)] \leq 0$$

$$\Rightarrow [8(1+\gamma_{21})|\psi_{2P}|^2 + 2\gamma_{21}(1+\Delta_1^2)][2(1+\gamma_{21})(1+\Delta_1^2 + 4\gamma_{21}) - 2\gamma_{21}(1+\Delta_1^2)]$$
$$- 4(1+\gamma_{21})^2[8|\psi_{2P}|^2(\gamma_{21} - \Delta_1\Delta_2)] + 16\gamma_{21}(1+\gamma_{21})(1+\Delta_1^2)|\psi_{2P}|^2$$
$$\leq 0$$

$$\Rightarrow [8(1+\gamma_{21})|\psi_{2P}|^2 + 2\gamma_{21}(1+\Delta_1^2)]\Big[2\Big(1 + \Delta_1^2 + 4\gamma_{21}(1+\gamma_{21})\Big)\Big]$$
$$- 4(1+\gamma_{21})^2[8|\psi_{2P}|^2(\gamma_{21} - \Delta_1\Delta_2)] + 16\gamma_{21}(1+\gamma_{21})(1+\Delta_1^2)|\psi_{2P}|^2$$
$$\leq 0$$

$$\Rightarrow \Big[16(1+\gamma_{21})(1+\Delta_1^2 + 4\gamma_{21} + 4\gamma_{21}^2)|\psi_{2P}|^2$$
$$+ 2\gamma_{21}(1+\Delta_1^2)\Big[2\Big(1 + \Delta_1^2 + 4\gamma_{21}(1+\gamma_{21})\Big)\Big]\Big] \quad (1.67)$$
$$- 16(1+\gamma_{21})^2(2\gamma_{21} - 2\Delta_1\Delta_2)|\bar{\psi}_{2P}|^2 + 16\gamma_{21}(1+\gamma_{21})(1+\Delta_1^2)|\psi_{2P}|^2$$
$$\leq 0$$

$$\Rightarrow \Big[16(1+\gamma_{21})(1+\Delta_1^2 + 4\gamma_{21} + 4\gamma_{21}^2)|\psi_{2P}|^2$$
$$+ 4\gamma_{21}(1+\Delta_1^2)\Big(1 + \Delta_1^2 + 4\gamma_{21}(1+\gamma_{21})\Big)\Big]$$
$$- 16(1+\gamma_{21})(1+\gamma_{21})(2\gamma_{21} - 2\Delta_1\Delta_2)|\psi_{2P}|^2$$
$$+ 16\gamma_{21}(1+\gamma_{21})(1+\Delta_1^2)|\psi_{2P}|^2 \leq 0$$

$$\Rightarrow \Big[16(1+\gamma_{21})(1+\Delta_1^2 + 4\gamma_{21} + 4\gamma_{21}^2)|\psi_{2P}|^2$$
$$+ 4\gamma_{21}(1+\Delta_1^2)\Big(1 + \Delta_1^2 + 4\gamma_{21}(1+\gamma_{21})\Big)\Big]$$
$$- 16(1+\gamma_{21})(1+\gamma_{21})(2\gamma_{21} - 2\Delta_1\Delta_2)|\psi_{2P}|^2$$
$$+ 16\gamma_{21}(1+\gamma_{21})(1+\Delta_1^2)|\psi_{2P}|^2 \leq 0$$

$$\Rightarrow \left[16(1+\gamma_{21})(1+\Delta_1^2+4\gamma_{21}+4\gamma_{21}^2)|\psi_{2P}|^2\right.$$
$$+ 4\gamma_{21}(1+\Delta_1^2)\left(1+\Delta_1^2+4\gamma_{21}(1+\gamma_{21})\right)\right]$$
$$- 16(1+\gamma_{21})(2\gamma_{21}-2\Delta_1\Delta_2+2\gamma_{21}^2-2\gamma_{21}\Delta_1\Delta_2)|\psi_{2P}|^2$$
$$+ 16\gamma_{21}(1+\gamma_{21})(1+\Delta_1^2)|\psi_{2P}|^2 \leq 0$$
$$\Rightarrow \left[16(1+\gamma_{21})^2(1+\Delta_1^2+2\gamma_{21}+2\Delta_1\Delta_2)|\psi_{2P}|^2\right.$$
$$+ 4\gamma_{21}(1+\Delta_1^2)\left(1+\Delta_1^2+4\gamma_{21}(1+\gamma_{21})\right)\right] \leq 0$$
$$\Rightarrow |\psi_{2P}|^2 \geq -\frac{\gamma_{21}(1+\Delta_1^2)\left(1+\Delta_1^2+4\gamma_{21}(1+\gamma_{21})\right)}{4(1+\gamma_{21})^2(1+\Delta_1^2+2\gamma_{21}+2\Delta_1\Delta_2)}$$

$|\psi_{2P}|^2$ can only be positive if $1+\Delta_1^2+2\gamma_{21}+2\Delta_1\Delta_2 \leq 0$, which in turn sets the frequency matching condition for phononic frequency combs.

## S2. Parameters relating to the existence boundaries of phononic frequency combs

### S2-1. Center frequency of existence band of phononic frequency combs

The frequency matching condition for phononic frequency combs is governed by eq. (2.1).
$$(1+\Delta_1^2+2\gamma+2\Delta_1\Delta_2) \leq 0 \tag{2.1}$$

Since $\Delta_2 = \frac{\Delta_1}{2} + \bar{\kappa}$, we have the following expression.

$$\left(1+\Delta_1^2+2\gamma+2\Delta_1\left(\frac{\Delta_1}{2}+\bar{\kappa}\right)\right) \leq 0 \tag{2.2}$$

$$\left(2\Delta_1^2+2\Delta_1\bar{\kappa}+(2\gamma+1)\right) \leq 0 \tag{2.3}$$

$$\frac{-\bar{\kappa}-\sqrt{\bar{\kappa}^2-2(2\gamma+1)}}{2} \leq \Delta_1 \leq \frac{-\bar{\kappa}+\sqrt{\bar{\kappa}^2-2(2\gamma+1)}}{2} \tag{2.4}$$

The center frequency of existence band of phononic frequency combs is calculated as follows.

$$\Delta_{1,center} = -\frac{\bar{\kappa}}{2}$$
$$\Rightarrow \frac{\omega_{D,center}-\omega_1}{\gamma_1} = -\frac{\omega_1-2\omega_2}{4\gamma_1} \tag{2.5}$$

$$\Rightarrow \omega_{D,center} = \frac{3\omega_1 + 2\omega_2}{4}$$

## S2-2. Range of existence band of phononic frequency combs

The range of existence band of phononic frequency combs is calculated as follows.

$$\Delta_{1,max} - \Delta_{1,min} = \sqrt{\bar{\kappa}^2 - 2(2\gamma + 1)}$$

$$\Rightarrow (\omega_{D,max} - \omega_{D,min})^2 = \gamma_1^2 \left( \frac{(\omega_1 - 2\omega_2)^2}{4\gamma_1^2} - \frac{2(2\gamma_2 + \gamma_1)}{\gamma_1} \right)$$

$$\Rightarrow R^2 = \left( \frac{(\omega_1 - 2\omega_2)^2 - 8(2\gamma_1\gamma_2 + \gamma_1^2)}{4} \right)$$

$$\Rightarrow R^2 = \left( \frac{\omega_1^2 - 4\omega_1\omega_2 + 4\omega_2^2 - 8\left(\frac{2\omega_1\omega_2}{Q_1 Q_2} + \frac{\omega_1^2}{4Q_1^2}\right)}{4} \right) \tag{2.6}$$

$$\Rightarrow R^2 = \left( \omega_2^2 - \left(1 + \frac{4}{Q_1 Q_2}\right)\omega_1\omega_2 + \frac{\omega_1^2}{4}\left(1 - \frac{2}{Q_1^2}\right) \right)$$

$$\Rightarrow R^2 = \left( \omega_2^2 - \omega_1\omega_2 + \frac{\omega_1^2}{4} \right) - \left( \frac{4\omega_1\omega_2}{Q_1 Q_2} \right) - \left( \frac{\omega_1^2}{2Q_1^2} \right)$$

For $Q_1^2 \gg 2$, $\frac{\omega_1^2}{4} \gg \frac{\omega_1^2}{2Q_1^2}$. Hence, $\frac{\omega_1^2}{2Q_1^2}$ can be neglected from eq. (2.6).

$$R^2 = \left( \omega_2^2 - \omega_1\omega_2 + \frac{\omega_1^2}{4} \right) - \left( \frac{4\omega_1\omega_2}{Q_1 Q_2} \right)$$

$$\Rightarrow R^2 = \left( \omega_2^2 - \omega_1\omega_2 \left(1 + \frac{4}{Q_1 Q_2}\right) + \frac{\omega_1^2}{4} \right) \tag{2.7}$$

The quadratic on $\omega_2$ in the RHS of eq. (2.7) can be written as

$$R^2 = \left( \omega_2 - \omega_1 \left[ \frac{\left(1 + \frac{4}{Q_1 Q_2}\right) + \sqrt{\left(1 + \frac{4}{Q_1 Q_2}\right)^2 - 1}}{2} \right] \right)^2$$

$$R^2 = \left( \omega_2 - \frac{\omega_1}{2} \left( \left(1 + \frac{4}{Q_1 Q_2}\right) + \sqrt{\frac{8}{Q_1 Q_2} + \frac{16}{Q_1^2 Q_2^2}} \right) \right)^2 \tag{2.8}$$

For $Q_1Q_2 \gg 2$, $\frac{8}{Q_1Q_2} \gg \frac{16}{Q_1^2Q_2^2}$. Hence, the term $\frac{16}{Q_1^2Q_2^2}$ can be neglected in eq. (2.8).

$$R^2 = \left(\omega_2 - \frac{\omega_1}{2}\left(1 + \frac{4}{Q_1Q_2} + \sqrt{\frac{8}{Q_1Q_2}}\right)\right)^2 \tag{2.9}$$

For $Q_1Q_2 \gg 2$, $\sqrt{\frac{8}{Q_1Q_2}} \gg \frac{4}{Q_1Q_2}$. Hence, the term $\frac{4}{Q_1Q_2}$ can be neglected in eq. (2.9).

$$R^2 = \left(\omega_2 - \frac{\omega_1}{2}\left(1 + \sqrt{\frac{8}{Q_1Q_2}}\right)\right)^2 \tag{2.10}$$

An expression for range can hence be obtained as follows.

$$R = \left|\omega_2 - \frac{\omega_1}{2} - \frac{\sqrt{2}\omega_1}{\sqrt{Q_1Q_2}}\right| \tag{2.11}$$

**S2-3. Minimum and maximum bound of existence band of phononic frequency combs**

The minimum bound of existence band of phononic frequency combs is calculated as follows.

$$\begin{aligned}\Delta_{1,min} &= \frac{-\bar{\kappa} - \sqrt{\bar{\kappa}^2 - 2(2\gamma + 1)}}{2}\\ \Rightarrow \frac{\omega_{D,min} - \omega_1}{\gamma_1} &= -\frac{\omega_1 - 2\omega_2}{4\gamma_1} - \frac{R}{2\gamma_1}\\ \Rightarrow \omega_{D,min} &= \omega_{D,center} - \frac{R}{2}\end{aligned} \tag{2.12}$$

The maximum bound of existence band of phononic frequency combs is calculated as follows.

$$\begin{aligned}\Delta_{1,max} &= \frac{-\bar{\kappa} + \sqrt{\bar{\kappa}^2 - 2(2\gamma + 1)}}{2}\\ \Rightarrow \frac{\omega_{D,max} - \omega_1}{\gamma_1} &= -\frac{\omega_1 - 2\omega_2}{4\gamma_1} + \frac{R}{2\gamma_1}\\ \Rightarrow \omega_{D,max} &= \omega_{D,center} + \frac{R}{2}\end{aligned} \tag{2.13}$$

## S2-4. Critical detuning for the generation of phononic frequency combs

For $\omega_2 - \frac{\omega_1}{2} < 0$, the right bound of regime of phononic frequency combs is closer to $\omega_1$, i.e., $\omega_{D,center} + \frac{R}{2}$.

$$\omega_{D,edge} = \omega_{D,center} + \frac{R}{2} = \frac{3\omega_1 + 2\omega_2}{4} - \frac{\omega_2}{2} + \frac{\omega_1}{4} + \frac{\omega_1}{\sqrt{2Q_1Q_2}}$$
$$\Rightarrow \omega_{D,edge} = \omega_{D,center} + \frac{R}{2} = \omega_1 + \frac{\omega_1}{\sqrt{2Q_1Q_2}}$$
(2.14)

For $\omega_2 - \frac{\omega_1}{2} > 0$, the left bound of regime of phononic frequency combs is closer to $\omega_1$, i.e., $\omega_{D,center} - \frac{R}{2}$.

$$\omega_{D,edge} = \omega_{D,center} - \frac{R}{2} = \frac{3\omega_1 + 2\omega_2}{4} - \frac{\omega_2}{2} + \frac{\omega_1}{4} + \frac{\omega_1}{\sqrt{2Q_1Q_2}}$$
$$\Rightarrow \omega_{D,edge} = \omega_{D,center} + \frac{R}{2} = \omega_1 + \frac{\omega_1}{\sqrt{2Q_1Q_2}}$$
(2.15)

Hence, the minimum detuning from $\omega_1$ that is required for exciting phononic combs can be obtained as

$$\delta = |\omega_{D,edge} - \omega_1| = \frac{\omega_1}{\sqrt{2Q_1Q_2}}$$
(2.16)

## S2-5. Critical separation of resonance frequencies $\left|\omega_2 - \frac{\omega_1}{2}\right|_c$ for the generation of phononic frequency combs

Since the condition for $\bar{\Delta}_1$ in eq. (2.4) can assume real values only if

$$\bar{\kappa}^2 \geq 2(2\gamma + 1)$$
$$\Rightarrow (\omega_1 - 2\omega_2)^2 \geq 8\gamma_1(2\gamma_2 + \gamma_1)$$
$$\Rightarrow \omega_1^2 - 4\omega_1\omega_2 + 4\omega_2^2 \geq \left(\frac{8\omega_1\omega_2}{Q_1Q_2} + \frac{4\omega_1^2}{Q_1^2}\right)$$
$$\Rightarrow \omega_2^2 - \left(1 + \frac{2}{Q_1Q_2}\right)\omega_1\omega_2 + \frac{\omega_1^2}{4}\left(1 - \frac{4}{Q_1^2}\right) \geq 0$$
$$\Rightarrow \omega_2 \geq \frac{\omega_1}{2}\left[\left(1 + \frac{2}{Q_1Q_2}\right) \pm \sqrt{\left(1 + \frac{2}{Q_1Q_2}\right)^2 - \left(1 - \frac{4}{Q_1^2}\right)}\right]$$
(2.17)

$$\Rightarrow \omega_2 \geq \frac{\omega_1}{2} + \frac{\omega_1}{Q_1 Q_2}\left(1 \pm 2\sqrt{Q_2^2 + Q_1 Q_2 + 1}\right)$$

For $Q_1 Q_2 \gg Q_2^2, 1$, we can neglect certain terms to obtain a simplified expression for the critical separation of resonance frequencies $g = \left|\omega_2 - \frac{\omega_1}{2}\right|_c$.

$$\left|\omega_2 - \frac{\omega_1}{2}\right| \geq 2\omega_1 \sqrt{\frac{1}{Q_1 Q_2}}$$
$$g = 2\omega_1 \sqrt{\frac{1}{Q_1 Q_2}}$$
(2.18)

### S2-6. Critical quality factor $Q_{2,c}$ for the generation of phononic frequency combs

An expression for the critical quality factor $Q_{2,c}$ can also be obtained as

$$\begin{aligned}
\bar{\kappa}^2 &\geq 2(2\gamma + 1) \\
(\omega_1 - 2\omega_2)^2 &\geq 8\gamma_1(2\gamma_2 + \gamma_1) \\
(\omega_1 - 2\omega_2)^2 &\geq \frac{8\omega_1}{2Q_1}\left(\frac{2\omega_2}{2Q_2} + \frac{\omega_1}{2Q_1}\right) \\
(\omega_1 - 2\omega_2)^2 &\geq \left(\frac{16\omega_1 \omega_2}{4Q_1 Q_2} + \frac{8\omega_1^2}{4Q_1^2}\right) \\
(\omega_1 - 2\omega_2)^2 - \frac{8\omega_1^2}{4Q_1^2} &\geq \frac{16\omega_1 \omega_2}{4Q_1 Q_2} \\
(\omega_1 - 2\omega_2)^2 - \frac{2\omega_1^2}{Q_1^2} &\geq \frac{4\omega_1 \omega_2}{Q_1 Q_2} \\
\frac{[(\omega_1 - 2\omega_2)^2 Q_1^2 - 2\omega_1^2]}{Q_1} &\geq \frac{4\omega_1 \omega_2}{Q_2} \\
Q_2 &\geq \frac{4\omega_1 \omega_2 Q_1}{[(\omega_1 - 2\omega_2)^2 Q_1^2 - 2\omega_1^2]}
\end{aligned}$$
(2.19)

For $Q_1 \gg 2$, the term $2\omega_1^2$ can be neglected in the denominator. Hence, we have

$$\begin{aligned}
Q_2 &\geq \frac{4\omega_1 \omega_2}{(\omega_1 - 2\omega_2)^2 Q_1} \\
Q_{2,c} &= \frac{4\omega_1 \omega_2}{(\omega_1 - 2\omega_2)^2 Q_1} \\
Q_{2,c} &= \frac{2}{\left(1 - \frac{\omega_1}{2\omega_2}\right)^2 Q_1}
\end{aligned}$$
(2.20)